\documentclass[twocolumn,amssymb,amsmath,preprintnumbers,floatfix,aps]{revtex4-1}
\usepackage{slashed,graphicx,bm,psfrag, amsmath, bbm,hyperref,epsfig,amssymb,euscript,latexsym,ulem}

\newcommand{\beq}{\begin{equation}}
\newcommand{\eeq}{\end{equation}}
\newcommand{\beqa}{\begin{eqnarray}}
\newcommand{\eeqa}{\end{eqnarray}}
\newcommand{\ba}{\begin{array}}
\newcommand{\ea}{\end{array}}

\def\CM {{\cal M}}
\def\CO {{\cal O}}

\newcommand{\be}{\begin{equation}}
\newcommand{\ee}{\end{equation}}
\newcommand{\bea}{\begin{eqnarray}}
\newcommand{\eea}{\end{eqnarray}}

\newcommand{\nn} {\nonumber\\}

\newcommand{\U}{\rm U}
\newcommand{\SU}{\mathop{\rm SU}}

\newcommand{\ex}[1]{{\rm e}^{#1}}

\def\a{\alpha}
\def\b{\beta}
\def\c{\chi}
\def\g{\gamma}
\def\C{\Gamma}
\def\d{\delta}
\def\D{\Delta}
\def\e{\epsilon}
\def\ve{\varepsilon}

\def\f{\phi}
\def\F{\Phi}
\def\p{\psi}
\def\P{\Psi}
\def\vf{\varphi}

\def\l{\lambda}

\def\m{\mu}
\def\n{\nu}
\def\r{\rho}
\def\s{\sigma}

\def\vth{\vartheta}

\def\o{\omega}

\def\z{\zeta}

\def\rmi{{\,{\rm i}\,}}

\def\del{\partial}
\def\Id{{\Bbb I}}
\def\rmd{{\rm d}}
\def\dsl{\slashed{\del}}

\def\tilde{\widetilde}
\def\bar{\overline}
\def\CO{\mathcal O}
\def\CN{\mathcal N}

\def\CM{\mathcal M}

\catcode`@=12

\numberwithin{equation}{section}

\begin{document}

\title{\boldmath Holographic Goldstino}

\author{Riccardo Argurio}
\affiliation{Physique Th\'eorique et Math\'ematique and International Solvay Institutes, Universit\'e Libre de Bruxelles, C.P. 231, 1050 Bruxelles, Belgium}

\author{Matteo Bertolini}

\affiliation{SISSA and INFN - Sezione di Trieste, Via Bonomea 265; I 34136 Trieste, Italy\\
ICTP, Strada Costiera 11; I 34014 Trieste, Italy}

\author{Daniele Musso}

\affiliation{ICTP, Strada Costiera 11; I 34014 Trieste, Italy}

\author{Flavio Porri}

\affiliation{Institute for Theoretical Physics and Center for Extreme Matter and Emergent Phenomena \\ 
Utrecht University, Leuvenlaan 4, 3584 CE Utrecht, The Netherlands}

\author{Diego Redigolo}

\affiliation{Sorbonne Universit\'es, UPMC Univ Paris 06 and CNRS \\ UMR 7589, LPTHE, F-75005, Paris, France}

\date{\today}

\begin{abstract}

\noindent
We find the fingerprints of the  
Goldstino associated to
spontaneous supersymmetry breaking in a prototype holographic model
for a strongly coupled  field theory. The Goldstino massless pole arises in
two-point correlators of the supercurrent, due to contact terms in supersymmetry 
Ward identities. We show how these
contact terms are obtained from the
holographic renormalization of the gravitino sector, independently of
the details of the bulk background solution. For completeness, we prove the existence of a family of such solutions in a simple supergravity model.

\end{abstract}

\maketitle

\section{Introduction}
The spontaneous breaking of global and space-time symmetries leaves universal fingerprints in the IR behavior of quantum field theories. In particular, the Goldstone theorem ensures that  massless modes should appear in the IR spectrum. In the case of spontaneously broken $\mathcal{N}=1$ rigid supersymmetry, we expect a massless Weyl fermion in the IR spectrum, the Goldstino.

The spontaneous breaking of $\mathcal{N}=1$ supersymmetry in 4d QFTs can be studied purely from the operator point of view \cite{Komargodski:2009rz}. In particular, the Goldstino appears as a massless pole in the supercurrent two-point correlator \cite{Argurio:2013uba}, while supersymmetry Ward identities relate the pole residue to the non-zero vacuum energy.

When the field theory dynamics is strongly coupled, its operators cannot be written in terms of fundamental fields and the theory becomes incalculable. In such a situation, the AdS/CFT correspondence \cite{Maldacena:1997re,Gubser:1998bc,Witten:1998qj} provides a powerful tool, allowing one to construct (at least in principle) supergravity backgrounds which are holographically dual to strongly coupled gauge theories. 

While much effort has been devoted to the study of supergravity backgrounds dual to strongly coupled QFTs which spontaneously break supersymmetry, very little has been done to understand how the holographic map can be extended to the supercurrent sector \footnote{Some related aspects have been studied at finite temperature in the
context of holographic models for strongly coupled fluids \cite{Policastro:2008cx,Gauntlett:2011mf,Gauntlett:2011wm,Kontoudi:2012mu,Erdmenger:2013thg}.}. This is particularly interesting because a holographic handle on the supercurrent correlators can be used to identify the appearance of the Goldstino mode from the bulk perspective. More generally, the supersymmetry Ward identities which constrain the boundary field theory should be encoded automatically in the dual supergravity theory.    

The primary objective of this paper is to fill the above mentioned gap in the literature, providing a holographic derivation of the Ward identities. This will allow us to identify the appearance of the Goldstino mode in the supercurrent two-point correlator. 

Previous works have already shown how the holographic renormalization procedure relates (broken) gauge symmetry/diffeomorphisms  in the bulk and (broken) flavor/conformal symmetry at the boundary \cite{Henningson:1998gx,Bianchi:2001de,Bianchi:2001kw}. Our work will complement these ideas by studying the connection between the (broken) bulk supersymmetry and the global supersymmetry of the boundary theory. 

In the remainder of this section we outline the general philosophy underlying our analysis. We elucidate general properties supersymmetric QFTs (SQFT) should enjoy, as well as those of the dual supergravity backgrounds. 

\subsection{Supersymmetry Ward identities and the Goldstino mode}
\label{intro1}

A necessary condition for a supersymmetric field theory to develop a Poincar\'e invariant supersymmetry-breaking vacuum is that conformality should be broken explicitly. 
Let us then focus on 4d SQFTs which can be described by RG flows departing from a UV fixed point by means of a relevant operator. 
Their action can be schematically written as 
\begin{equation}
\label{defqft}
S  = S _{SCFT} + \lambda \int\rmd^4 x \,\rmd^2\theta \, \CO + h.c.~,
\end{equation} 
where $ S _{SCFT}$ is an ${\cal N}=1$ superconformal field theory and $\CO$  is a chiral operator with $1\leq\Delta_\CO < 3$ which triggers the RG-flow \footnote{As shown in \cite{Green:2010da} the superpotential deformation in \eqref{defqft} is actually the only possible relevant deformation one can write.}. Its operator components are specified as 
$\CO = \CO_S+  \sqrt{2}\, \theta \, \CO_\psi + \theta^2 \CO_F$. In principle, one can deform the SCFT with more than one relevant operator. We make the simplifying assumption of having only one such operator.

In any SQFT, among all the supermultiplets of gauge invariant operators that one can construct, 
a prominent role is played by the supercurrent multiplet which contains the energy-momentum tensor $T_{\mu\nu}$ 
and the supersymmetry current $S_{\mu\alpha}$. 
For non-conformal theories, as the one under consideration, this multiplet is 
described \footnote{In what follows we assume that a Ferrara-Zumino (FZ) multiplet can always be defined. See \cite{Komargodski:2010rb,Dumitrescu:2011iu} for a discussion of the physical requirements underlying this assumption.}
by two superfields $({\cal J}_\mu, X)$ satisfying the on-shell relation 
\begin{align}
\label{fzmul}
- 2 \bar D^{\dot\alpha} \sigma^\mu_{\alpha\dot\alpha} {\cal J}_\mu = D_\alpha X~,
\end{align}
with ${\cal J}_\mu$ a real superfield and $X$ a chiral superfield. The latter contains trace operators, in particular those of the energy-momentum tensor and the supercurrent, and vanishes for a SCFT (we refer to \cite{Argurio:2013uba} and \cite{Komargodski:2010rb} for details). In \eqref{defqft} $\CO$ is related to $X$ as
\begin{equation}\label{XPhi}
X = \frac 43 (3 - \Delta_\CO) \lambda \, \CO ~.
\end{equation}
From here on we take $\Delta_\CO = 2$ and hence $X = \frac 43 \,\lambda \,\CO$. Nothing relevant of what we discuss below depends on this choice. Notice that eq.~\eqref{XPhi}, being an operator identity, holds inside any correlation function and in any vacuum.

The theory can be in a supersymmetric or a supersymmetry breaking vacuum, depending on whether the operator $\CO$ (or any other operator of the strongly coupled field theory) acquires a non-vanishing VEV for its F-term component. The structure of one and two-point functions of operators belonging to the FZ multiplet  can easily tell if this is the case. Indeed, regardless of the vacuum one is considering, the supersymmetry algebra implies the Ward identities
\begin{align}
\label{wiS}
& \langle \partial^\mu S_{\mu\alpha}(x) \,\bar S_{\nu\dot\beta}(0) \rangle = -\d^4(x) \langle \delta_\alpha \bar S_{\nu\dot\beta} \rangle =  - 2  \sigma^\mu_{\ \alpha\dot\beta} \, \langle T_{\mu\nu}\rangle \, \d^4(x) \\ 
\label{wipsi}
& \langle \partial^\mu S_{\mu\alpha}(x) \,  {\CO_\psi}_{\beta}(0)\rangle = -\d^4(x) \langle \delta_\alpha  {\CO_\psi}_{\beta} \rangle =  \sqrt{2} \,\langle {\cal O}_F \rangle \, \varepsilon_{\alpha\beta} \, \d^4(x)\ ,
\end{align}
where $\langle T_{\mu\nu}\rangle$ is a function of both $\lambda$ and $\langle {\CO_F} \rangle$ that vanishes if either of the two does (from eq.~\eqref{XPhi} it follows that $ \eta^{\mu\nu}\langle T_{\mu\nu} \rangle = 2\l {\rm Re}\langle \CO_F \rangle$). 
The two Ward identities above imply the presence of the following structures in the two-point functions of the supercurrent with itself and with the fermionic operator $\CO_\psi$
\begin{align}
\label{SS}
& \langle S_{\mu\alpha}(x)\, \bar{S}_{\nu\dot\beta}(0)\rangle = \dots - \frac{\rmi}{4\pi^2}\langle T\rangle\, (\sigma_\mu \bar\sigma^\rho  \sigma_\nu)_{\alpha\dot\beta}
\frac{x_\r}{x^4}\\
\label{psipsi}
&  
\langle S_{\mu\alpha}(x) \, {\CO_\psi}_{\beta}(0)\rangle = \dots  - \frac\rmi{2\pi^2}   \sqrt{2}\, \langle {\CO_F} \rangle \, \varepsilon_{\a\b}
\frac{x_\m}{x^4}\,,
\end{align}
where $\langle T\rangle= \eta^{\mu\nu}\langle T_{\mu\nu}\rangle$. 
Upon Fourier transform, the expressions above display in the simplest and cleanest way the massless pole associated to the Goldstino, which is the
lowest energy excitation in both $S_\mu$ and $\CO_\psi$. Indeed, in the deep IR, one can write 
$S_\mu=\sigma_\mu \bar{G}$, where $G$ is the Goldstino field. 
Plugging this relation in \eqref{SS}, one recovers, up to an overall normalization, the Goldstino propagator.

Finally, eq.~\eqref{XPhi} implies also that $\sigma^\mu_{\ \alpha \dot\a}\, \bar S^{\dot\a}_\mu = 2\sqrt2\, \l \, {\CO_\psi}_\a $, which in turn provides identities between a priori different correlation functions, e.g. 
\begin{equation}
\label{opipsi}
\langle \sigma^\mu_{\ \alpha \dot\a}\, \bar S^{\dot\a}_\mu(x) \,  {\bar\CO_\psi}_{\dot\b}(0) \rangle = 2\sqrt2\, \l \, \langle {\CO_\psi}_\a(x) \, {\bar\CO_\psi}_{\dot\b}(0) \rangle ~.
\end{equation}

\subsection{Field/Operator map}
\label{intro2}

The quantum field theory \eqref{defqft} can be holographically described with a five-dimensional ${\cal N} = 2$ supergravity theory containing just one hypermultiplet besides the always present graviton multiplet. The former contains a Dirac hyperino and two complex scalars, $\r$ and $\f$. The role of the hypermultiplet is twofold. First, some non-trivial scalar profile is needed in order to describe
holographically a non-conformal QFT. 
Second, from the field/operator map, one easily understands that the degrees of freedom of the hypermultiplet are needed to match those of the FZ multiplet whenever $X \not = 0$ \cite{Argurio:2013uba}. Indeed, the hypermultiplet is dual to the operator $\CO$ which, in turn, is related to $X$. 

Since $\CO$ is a relevant operator, the dual backgrounds are Asymptotically AdS (AAdS), 
meaning that we can use, to a large extent,  standard AdS/CFT techniques \cite{Girardello:1998pd,Girardello:1999bd,Freedman:1999gp}. 
In particular, we can use the well-known formula
\begin{equation}
\label{dimsc}
m^2 =\D(\D-4)~,
\end{equation}
which relates the (AdS) mass of a scalar field with the dimension of the dual QFT operator. 
Since $\Delta(\CO)=2$, the dimensions of the two scalar operators are $\Delta({\cal O}_S)=2$ and $\D({\CO_F})=3$. 
This implies, from eq.~\eqref{dimsc}, that  the two hyperscalars should have $m^2=-4,-3$, with the following field/operator map
\begin{equation}
\label{dualopsc}
\r \longleftrightarrow {\cal O}_S\quad , \quad \f \longleftrightarrow {\CO_F}~.
\end{equation}
Working in a coordinate system where the AAdS metric is
\begin{equation}
ds^2 = \frac{1}{z^2}\left(F(z) dx^2 + dz^2 \right) \quad , \quad  F(0) = 1 ~,\label{domainwall}
\end{equation}
and the AdS boundary is at $z=0$, the near boundary expansion for the two scalar fields is
\begin{equation}
\label{eqfc}
\f \sim z ( a + b z^2) + { O}(z^5)\quad , \quad
\r \sim z^2 ( c\log z + d ) + { O}(z^4) ~,
\end{equation}
where we take $a,b,c,d$ to be independent of the 4d coordinates in order to preserve Poincar\'e invariance at the boundary. Since the equations of motion for the scalars are second order,  a given choice of the leading and the subleading modes in the near boundary expansion determines the scalar profiles univocally. Switching on each of these modes corresponds to a specific deformation of the dual field theory which we classify here below.
\begin{itemize}
\item $a$ corresponds to a source for the operator ${\CO_F}$ and is related to the QFT coupling $\lambda$. As such, it should always be non-vanishing in order to describe the setup of \eqref{defqft}; 
\item $b$ is related to the VEV of ${\CO_F}$ and hence to the spontaneous breaking of supersymmetry; 
\item $c$ is a source for the QFT operator ${\cal O}_S$ and hence corresponds to a coupling which explicitly breaks supersymmetry. This soft breaking term is not present in \eqref{defqft} and should then be put to zero; 
\item $d$ is related to the supersymmetry preserving VEV of ${\cal O}_S$ and can be non-vanishing in any (either supersymmetry preserving or supersymmetry breaking) vacuum. 
\end{itemize}

In the backgrounds we consider, in which supersymmetry is either preserved or spontaneously broken, the scalar $\r$ should then have a vanishing value for the leading mode $c$. In fact, without affecting any of the main aspects we want to discuss, we could (and will) reduce to backgrounds where also $d=0$, and hence $\r=0$ altogether. In such single-scalar backgrounds, the difference between supersymmetric and non-supersymmetric vacua will then depend on the value of the subleading mode $b$. 

\vskip 10pt
In the remainder of this paper we put the general ideas outlined above in a concrete setting. In Section \ref{sugram} we present the simple gauged supergravity model we focus on, and discuss its corresponding supersymmetric and supersymmetry breaking solutions. In Sections \ref{renbos} and \ref{renfer} we perform holographic renormalization for the on-shell supergravity boundary action, a necessary step in order to compute holographically the correlators \eqref{wiS}-\eqref{opipsi}. Section \ref{golhol} contains the key results of our paper. In particular, using the well-known AdS/CFT prescription \cite{Witten:1998qj, Gubser:1998bc}, we derive equations \eqref{wiS}-\eqref{opipsi} holographically. We conclude in Section \ref{concl} with a summary of our results and an outlook.  Two Appendices contain technical details that we have omitted from the main text.

\section{The supergravity model}
\label{sugram}

The model we consider is a simplified version of the one studied for example in \cite{Ceresole:2001wi}. This is $\CN = 2$ 5d supergravity coupled to one hypermultiplet, with scalar 
manifold ${\cal M}=\SU(2,1)/(\U(1)\times\SU(2))$ and with the graviphoton gauging a proper $\U(1)$ subgroup of the isometries of ${\cal M}$.  The gauging, which determines the scalar potential and in turn the scalar masses, is fixed according to our choice for the dimension of the dual operator $\CO$. In fact, there exists a one parameter family of possible gaugings, which would allow to describe deformations by operators of any dimension. We refer to Appendix \ref{sugraapp} for further details on the model.

As anticipated, we want to focus on backgrounds with a single scalar having a non-trivial profile.\footnote{We can always choose it to be real without loss of generality.} We then start with the action
\begin{align}
\label{5dactionbos}
& S_\text{5D} = \int \rmd^5 x\; \sqrt{-G}\; \bigg\{ \,\frac12 R   -\del_M \f\, \del^M \f 	- U(\f)	  \bigg\} \,, 
\end{align}
which is obtained setting to zero all the fields but the metric and $\phi$ in \eqref{5daction} and where the scalar potential for $\f$ is given by
\begin{equation}
\label{pot}
U(\f) = \frac1{12} \left( 10 - \cosh(2\f) \right)^2 - \frac{51}4  ~.
\end{equation}
We look for solutions of the model \eqref{5dactionbos} taking the flat domain wall ansatz \eqref{domainwall} and requiring $\phi$ to depend only on the holographic coordinate $z$. Within these assumptions, the equations of motion for the bosonic sector read
\begin{subequations}\label{niceom}
\begin{align}
&6\left(1-\frac{zF'}{2F}\right)^2 = z^2\f'^2- U(\f)	\label{Eeq1}\\
&\left(1-\frac{zF'}{2F}\right)' = \frac23 z \f'^2		\label{Eeq2}\\
& z^2\f''-\left(3-2\frac{zF'}{F}\right)z\f' = \frac12 \del_\f U(\f) \,,\label{phieom}
\end{align}
\end{subequations}
where  $'$ denotes derivatives with respect to $z$. Equation \eqref{Eeq2} being redundant, a generic solution of the system \eqref{niceom} is fixed by three integration constants. One of these constants is fixed by the normalisation of the metric in \eqref{domainwall} while the other two are conveniently chosen to be $a$ and $b$ (i.e. the boundary conditions for the leading and the subleading mode of $\phi$ in \eqref{eqfc}).

Supersymmetric solutions must satisfy the following BPS system of first order differential equations\footnote{As usual, BPS equations are obtained as necessary conditions for the vanishing of the supergravity variations of the fermionic fields.}
\begin{subequations}\label{BPSsystem}
\begin{align}
& 1-\frac{zF'}{2F} =  W(\f) \\
& z\f' = \frac32\del_\f W(\f)\, ,
\end{align}
\end{subequations}
where the superpotential $W$ is given by
\begin{equation}
\label{sup}
 W(\f) = \frac16 \left( 5 + \cosh(2\f) \right)\ ,
\end{equation}
and satisfies the relation \eqref{reluw}. One can easily verify that the BPS system above implies the equations of motion \eqref{niceom}. Since the order of the equation of motion for $\phi$ is reduced, we expect a supersymmetric solution to be realized only for a specific relation between $a$ and $b$.

One supersymmetric solution of the system \eqref{BPSsystem} is just pure AdS, with $F_\text{AdS} = 1$ and $a=b=0$. Around this AdS solution the scalar mass is
\begin{align}\label{adsmassesbos}
 m^2_\f = \frac12 \del^2_\f U(0) = -3 \, ,
\end{align}
in units of the AdS radius.
This  shows that  the scalar field $\f$ is indeed suitable to be dual to an operator of dimension 3, such as $\CO_F$. 

Besides the pure AdS solution, the BPS system has other $z$-dependent solutions. Their general form, which can be found analytically, depends on the choice of one integration constant and reads
\begin{align}\label{susybkgsol}
\f(z) & = \frac12 \log\left( \frac{1+ a z}{1- a z }\right) \nn
F(z) & = \left( 1 -  a^2 z^2 \right)^{1/3} ~.
\end{align}
Comparing with \eqref{eqfc} one sees that $b = b_{susy}=a^3/3$. 
The solution \eqref{susybkgsol} represents a supersymmetric vacuum of the theory \eqref{defqft}, where $a$ will be identified with the coupling $\lambda$. The pure AdS solution is recovered for $a=0$. 

Let us now turn to the analysis of the second order equations of motion. The system \eqref{niceom} cannot be solved analytically, nonetheless it can be easily integrated numerically. 
The general solution depends on two parameters and its expression for small $z$ is given by the expansions
\begin{align}\label{bkgsol}
\f(z) & = a \,z+b \,z^3 +  O(z^5) \nn
F(z) & = 1 - \frac{a^2}3\,z^2 + \frac{a^4 - 9 a b}{18} \,z^4 + O(z^6)\,,
\end{align}
which reduce to the BPS case for $b = b_{susy}$. Conversely, for different values of $b$, the solutions are non-supersymmetric. Therefore, from here on we define the supersymmetry breaking order parameter as $\b = \frac{a^3}3 - b$. This will discriminate supersymmetric solutions, $\b=0$, from non-supersymmetric ones, $\b\neq0$. Recalling the discussion in Section \ref{intro2}, we then expect the VEV of the operator ${\cal O}_F$ to be proportional to $\b$.

\begin{figure}
\begin{center}
\includegraphics[height=0.16\textheight]{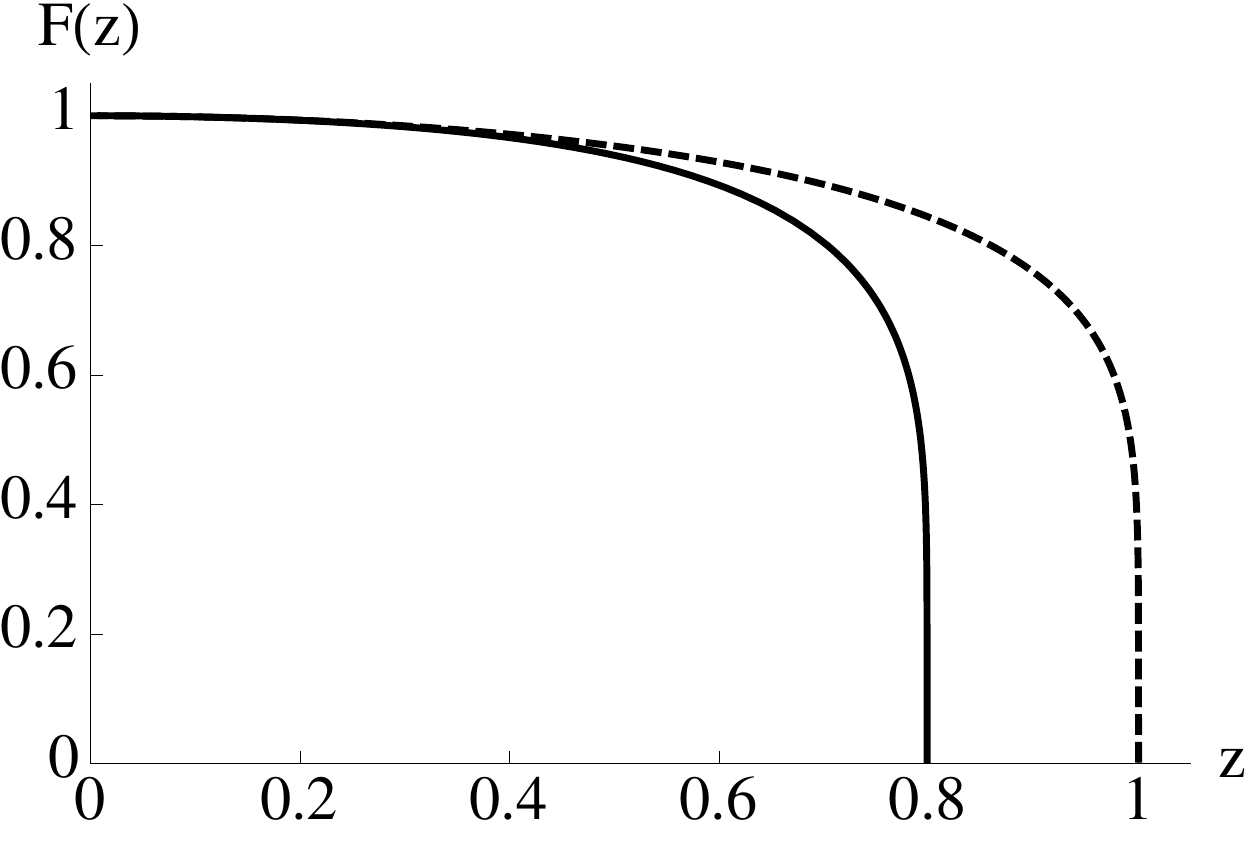}\hfill
\includegraphics[height=0.16\textheight]{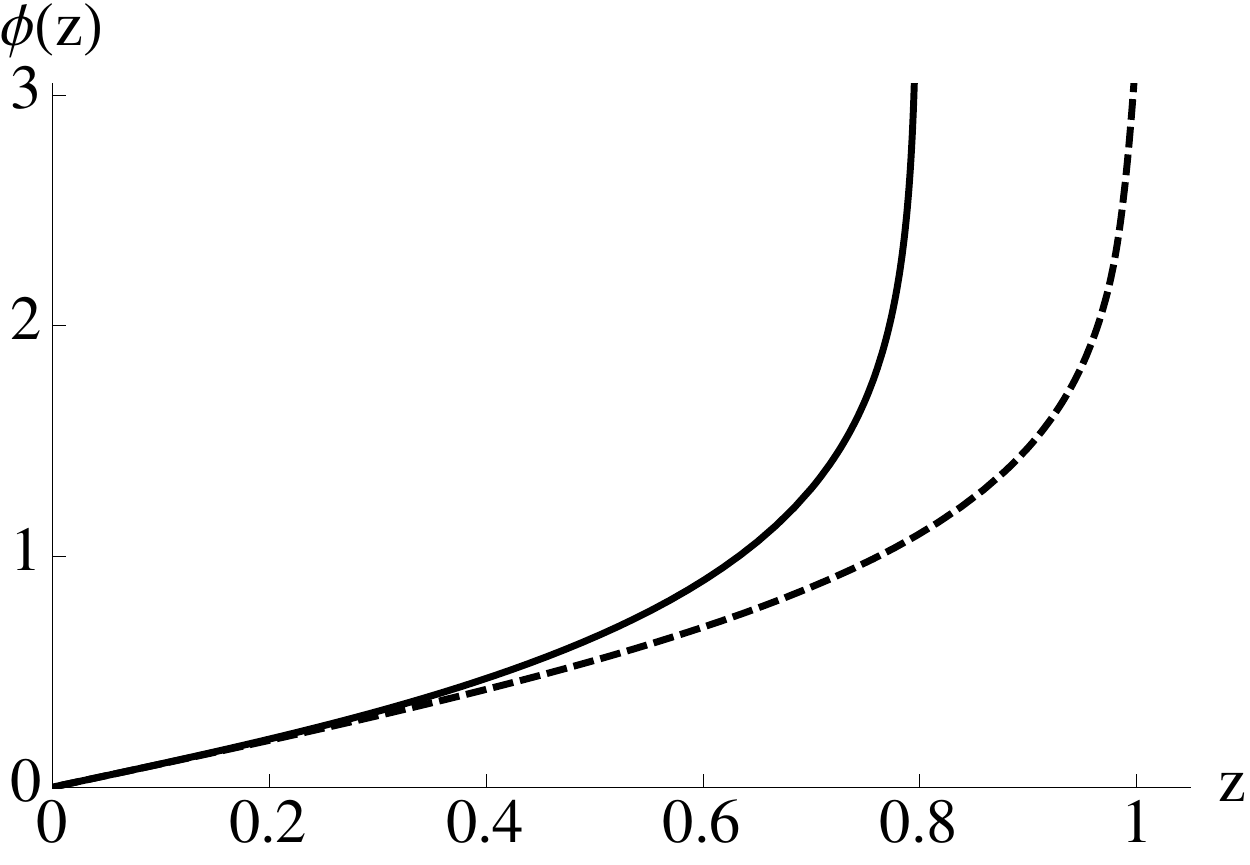}
\caption{\small  $F(z)$ and $\phi(z)$ profiles for a BPS solution (dashed) with $a=1$ and $\b=0$, and a supersymmetry-breaking one (solid) with $a=1$ and $\b = - 2/3$. While the behaviour is the same, switching on $\b$ the position of the singularity moves closer to the $z=0$ boundary. 
\label{susy1}}
\end{center}
\end{figure}

In Figure \ref{susy1} we show the profiles of the warp factor $F$ and the scalar $\phi$ for supersymmetric and non-supersymmetric solutions. 
They are both singular and in fact approach the singularity in a very similar way. These solutions are presented merely as an existence proof, and in the following we will not need to discuss their properties in any detail. In particular, the nature of the singularity does not affect our final results.

To wrap-up, we see that the model presented here is in fact a concrete example of the general picture outlined in Sections \ref{intro1} and \ref{intro2}. The scalar $\f$ is dual to a relevant operator of dimension 3 which triggers a non-trivial RG-flow out of some given UV SCFT fixed point. The solutions  \eqref{bkgsol} represent such non-trivial RG-flows. The dual QFT can find itself in a supersymmetric vacuum, $\langle {\CO_F}\rangle=0$, or a non-supersymmetric one $\langle {\CO_F} \rangle\neq0$. Likewise, the background solution can preserve bulk supersymmetry, $\b=0$, or break it, $\b\neq0$. One is then led, as already stressed, to identify $\b$ with the VEV of the QFT operator ${\CO_F}$. In what follows we will prove this to be indeed the case by a direct holographic computation.

\section{Holographic renormalization: bosonic sector}
\label{renbos}

In order to derive holographically the identities \eqref{wiS}-\eqref{wipsi}, we have to compute the one-point functions of the energy-momentum tensor and of $\CO_F$, which appear on the right hand side of eqs.~\eqref{wiS} and \eqref{wipsi}, respectively. This is done by applying standard holographic renormalization techniques to the coupled system of gravitational degrees of freedom and the scalar $\f$. 

Let us then consider the fluctuations of both the scalar $\f$ and the metric around their background values \eqref{bkgsol} 
\begin{align}
\phi  &=\phi(z)+\varphi(z,x)\nn 
ds^2  &=\frac{1}{z^2}\left(dz^2+F(z)\left(\eta_{\mu\nu}+h_{\mu\nu}(z,x)\right)dx^{\mu}dx^{\nu}\right)\ ,
\end{align}
where we fix the gauge $h_{zz}=h_{z\m}=0$. We now have to evaluate the on-shell action at the boundary $z=0$, and differentiate it with respect to the corresponding sources. For later convenience we decompose the 4d metric as
\begin{equation}
\label{comph}
h_{\mu\nu}=h_{\mu\nu}^{tt}+\eta_{\mu\nu}h+ \partial_{(\mu} H_{\nu)}\ .
\end{equation}
Since our goal here is just to compute one-point functions, we focus on the part of the on-shell action which is linear in the fluctuations, namely
\begin{equation}\label{bdyboson}
S^\text{B}_\text{bdy}= \int d^4x \sqrt{-g}   \left[\left(3-\frac{3zF'}{2F}\right)\left(1+2h+\dots\right) +2 z \, \phi'\varphi\right]~,
\end{equation}
where the $\dots$ stand for $H_\nu$ contributions, which are not relevant for what we do next. 
Evaluating the action at the boundary $z=0$, we get (as usual in AdS/CFT)  a divergent result. Hence, following the standard holographic renormalization procedure \cite{Bianchi:2001kw}, we regularize it at $z=\e$ and add covariant counterterms to subtract the divergent pieces. Such procedure, however, contains ambiguities in the form of finite counterterms that one can arbitrarily add. In our case, the natural choice is a supersymmetric subtraction scheme.
This can be enforced requiring the renormalized on-shell action to vanish on supersymmetric configurations. In this way we fix completely the set of counterterms we need, both divergent and finite.

As it is well known \cite{Bianchi:2001de}, the counterterm action in a supersymmetric scheme always contains a term proportional to the superpotential 
\begin{equation}\label{Wct}
S^\text{B}_\text{c.t.} = -\int d^4x \sqrt{-g}\, 3 \,W(\f + \vf)~,
\end{equation}
with $W$ defined in eq.~\eqref{sup}. Such counterterm does not remove all divergences for a general solution with $x$-dependent boundary conditions. However, in our case, at linear order in the fluctuations the counterterm action \eqref{Wct} turns out to reabsorb all divergences coming from \eqref{bdyboson}. 

The final renormalized action then reads
\begin{align}
\label{SrenTfin}
S^\text{B}_\text{ren}  &= \lim_{\e\to 0}  \left( S^\text{B}_\text{bdy}  + S^\text{B}_\text{c.t.} \right) \\
&= -\int d^4 x \; \left[a\b \left(1+2h_0+\dots \right)+4\b\varphi_0\right]~,
\end{align}
where $h_0$ and $\varphi_0$ are the leading boundary modes of the fields $h$ and $\varphi$, respectively. 
Notice that this action vanishes when the background is supersymmetric, $\b=0$. Furthermore, as expected, the coefficient of the metric fluctuation $h_0$ is proportional to both the background source $a$ and the supersymmetry breaking VEV $\b$.

\section{Holographic renormalization: fermionic sector}
\label{renfer}

Let us now consider the fermionic sector and the corresponding equations of motion for the gravitino and the hyperino. 
Within the gauging we have chosen, the fermionic part of the supergravity action is
\begin{align}
& S^\text{F}_\text{5D} = \int \rmd^5 x \sqrt{-G} \bigg\{ -  \bar\P_M\C^{MNP}D_N\P_P - 2\, \bar\z\C^M D_M\z 
\nn & + \rmi \del_N \f \left(  \bar\z \C^M\C^N\P_M \,  - \bar\P_M\C^N\C^M\z \right) \nn & +  2\,\CN(\f) \left( \bar\P_M\C^M\z + \bar\z\C^M\P_M \right) \nn 
 & + m(\f)\, \bar\P_M \C^{MN}\P_N - 2\,\CM(\f)\, \bar\z\z  \bigg\} \,, 
\label{Lferm2}
\end{align}
where $\CN$, $m$ and $\CM$ are defined in \eqref{fermmass}, and it is understood that $\f$ and the metric are set to their background values \eqref{bkgsol}. 
We observe that, around the AdS solution, we have
\begin{align}\label{adsmasses}
m_\P \equiv m(0) =  \frac32 \quad , \quad m_\z \equiv \CM(0) = -\frac12 \quad , \quad \CN(0)=0 \,.
\end{align}
The equations of motion derived from the action \eqref{Lferm2} are
\begin{subequations}\label{gravhypino}
\begin{align}
&\label{hypino}
      \C^M  D_M \z + \CM\, \z
      - \frac\rmi2 \C^M \C^N \P_M \,\del_N \f
      - \CN\, \C^M \P_M = 0 \\
&\label{gravino}
\C^{MNR} D_N \P_R - m \, \C^{MN} \P_N + \rmi \, \C^N\C^M \z \,\del_N \f \nn & - 2 \,\CN \,\C^M \z  = 0 \,.
\end{align}
\end{subequations}
In fact, the bulk action \eqref{Lferm2} has to be supplemented by boundary terms in order for it to be stationary on the equations of motion. Eventually the on-shell action reduces to the following boundary term\footnote{To the best of our knowledge the gravitino boundary action has been derived by \cite{Volovich:1998tj} in the pure AdS case. In Appendix \ref{bdyferm} we provide a derivation for AAdS backgrounds.}
\begin{align}
\label{bdygravhyp}
S^\text{F}_\text{bdy} = \int_{\del\CM} \rmd^4 x \sqrt{-g} \left\{ - \frac12 \bar\P_m \C^{mn} \P_n - \bar\z\z \right\}  ~,
\end{align}
where $g$ is the determinant of the induced metric on the boundary, $g_{mn} = \frac{F}{z^2} \eta_{mn}$. From here on we split the 5d curved indices as $M=m,z$, using latin alphabet for 4d curved indices and the letter $z$ to indicate indices along the fifth direction (we reserve $\mu$ for 4d flat indices). We will also work in the axial gauge, $\P_z=0$, which is the natural choice in a holographic setup. The 4d components of the gravitino will be further split as
\begin{equation}\label{gravexp}
\P_m = \p_m + \del_m \vth + \C_m \c \,, \qquad \del^m \p_m = \C^m \p_m = 0\,.
\end{equation}
Accordingly, the equations of motion \eqref{gravhypino} break down into
a transverse traceless part, which is decoupled (slashed quantities
are contracted with the curved $\C^m$, while
$\g^5\equiv\frac{1}{z}\C^z$ is the flat one)
\begin{equation}
\label{gravitinott}
\left(   z\del_z -1+\frac{zF'}{2F} + m\g^5 +\g^5\dsl\right) \p_m = 0\,,
\end{equation}
and a coupled system of equations for the longitudinal and traceful
parts of the gravitino and the hyperino
\begin{subequations} 
\label{gravitinoll}
\begin{align}
&  \left( z \partial_z -2+\frac{zF'}{F} + \mathcal{M} \gamma^5  + \gamma^5 \dsl \right) \zeta
         \nn & + \frac{1}{2 } \left( \rmi z \, \phi^\prime - 2 \,\mathcal{N} \gamma^5 \right) (\dsl\vth+4\c)      = 0     \\
& 3\g^5\dsl\c +   \left( -\frac{3}{2}  +\frac{3zF'}{4F}  - m \g^5  \right) (\dsl\vth+4\c) \nn &+  \left( \rmi z \, \phi^\prime  - 2 \,\mathcal{N} \gamma^5 \right)  \zeta =0 \\
&  \left(  3 z\del_z  -  3+\frac{ 3zF'}{2F} - m \g^5\right)(\dsl\vth+4\c) \nn & - 2\left(3 \rmi z\,\f' + 2\,\mathcal{N}\g^5 \right)\z  = 0 \\
& 3 \left(   z\del_z -  3+\frac{ 3zF'}{2F} + m \g^5\right) \dsl\c
- \left( \rmi z\,\f'- 2\,\mathcal N \g^5 \right) \dsl \z = 0 \,.
\end{align}
\end{subequations}
Provided one uses the equations of motion of the background \eqref{niceom}, it is
possible to show that only three of the above equations are independent. It is also worth noticing that the system of equations of motion above is invariant under the local transformations
\begin{align}\label{sugratr}
\d\vth &= \e \nn
\d\c &= \frac13\left[ m(\f) +\frac32 \left(\frac{z\,F'}{2F} - 1\right) \gamma^5 \right]\e \\
\d\z & = \left( \CN(\f)+\frac\rmi2 z\g^5\f' \right) \e \,, \notag
\end{align}
provided, again, that $\f$ and $F$ are restricted to their background values (i.e., they are solution to \eqref{niceom}), and that the local spinor parameter $\e(z,x)$ satisfies the differential equation
\begin{equation}\label{axgauge}
\left(z\del_z+ \frac{m}3\g^5\right)\e=0\,.
\end{equation}
The transformations \eqref{sugratr} are the local supersymmerty transformations of the supergravity theory restricted to a domain-wall background. The requirement \eqref{axgauge} comes from the axial gauge-fixing.

The near boundary expansions for the components of the gravitino field, up to the order we will need to write the on-shell action, are
\begin{align}
\label{relfer1}
&\begin{cases}
&\p^+_m = z^{-1/2} \left( \p_{m0}^+ + \p_{m2}^+\, z^2 \right) + O(z^{7/2})	\\
&\p^-_m = z^{1/2}   \left( \p_{m1}^- + \p_{m3}^- \,z^2\log\,z + \tilde \p_{m3}^-\, z^2 \right) + O(z^{9/2})
\end{cases}\\
\label{relfer2}
&\begin{cases}
&\vth^+ =  z^{-1/2} \left(  \vth_0^+ + \vth_2^+ \,z^2 \right) + O(z^{7/2}) \\
&\vth^- =    z^{1/2} \left( \vth_1^- +  \vth_3^- \,z^2\log\,z + \tilde \vth_3^- \,z^2 \right) + O(z^{9/2})
\end{cases}\\
\label{relfer3}
&\begin{cases}
&\c^+ = z^{7/2} \,\tilde \c_3^+ + O(z^{11/2})		\\
&\c^- =  z^{1/2} \left( \c_0^- + \c_2^- \,z^2 \right) + O(z^{9/2})
\end{cases} 
\end{align}
while for the hyperino 
\begin{align}
\label{relfer4}
&\begin{cases}
&\zeta^+ = z^{5/2} \left( \zeta_1^+ \,\log\,z +\tilde \zeta_1^+ \right) + O(z^{9/2})	\\
&\zeta^- = z^{3/2} \,\zeta_0^- + O(z^{7/2})	~,
\end{cases}
\end{align}
where the $\pm$ superscript denotes that the spinor is an eigenstate of $\g^5$ with eigenvalue $\pm 1$, i.e. $\g^5 \F^\pm = \pm \F^\pm$. 
 
Using the equations of motion \eqref{gravitinott}--\eqref{gravitinoll} we get the following
relations among the coefficients of the near-boundary expansions. The
transverse coefficients are given by
\begin{align}
\label{nearbdypsi}
& \p_{m1}^- = -\frac 12 \,\dsl\, \p_{m0}^+ \quad , \quad \p_{m3}^- = -\frac {1}{12}
\left(4 a^2 - 3\Box \right) 
\dsl \p_{m0}^+ \nn 
& \p_{m2}^+ = -\frac1{12}
\left( a^2 - 3\Box \right) \p_{m0}^+\ .
\end{align}
The longitudinal coefficients separate into the untilded ones, which
do not depend on the supersymmetry breaking parameter $\b$
\begin{align}
\label{nearbdychi}
& \zeta_1^+ = - \rmi a\,\dsl\, \c_0^- - \dsl\, \zeta_0^- \,, \, \vth_1^- = \c_0^-
 \, , \, \vth_3^- = \frac23 a^2\, \c_0^- - \frac23 a\rmi  \zeta_0^- \,,
 \nn 
& \vth_2^+ = -\frac{a^2}{12} \vth_0^+   \,,\,  \c_2^- = \frac{a^2}{12} \,\c_0^- + \frac{a}3\rmi  \zeta_0^- \,,
\end{align}
and the tilded ones which do depend on $\b$
\begin{align}
\label{nearbdychibeta}
&\tilde \vth_3^- = - \frac23 \beta \left(a \frac\dsl{\Box}\vth_0^+ + \frac{4 a}{\Box}\c_0^- -\rmi \frac{4}{\Box}\zeta_0^-\right) \nn &
  - \frac{a^2}4 \c_0^- + \rmi \frac{a}3  \zeta_0^- + \rmi \frac 23 a \frac{\dsl}{\Box} \,\tilde \zeta_1^+\nn 
&  \tilde \c_3^+ = \frac{\beta}{6} \left( a \, \vth_0^+ + 4\, a \frac{\dsl}{\Box}\c_0^- - \rmi 4 \frac\dsl{\Box} \zeta_0^- \right)\,.
\end{align}
This leaves us with six independent
coefficients: two transverse, $\p_{m0}^+,\, \tilde\p_{m3}^-$, and four
longitudinal, $\vth_0^+,\, \c_0^-\,, \z_0^-,\, \tilde\z_1^+$.

In order to compute correlators in the dual QFT, we have to
evaluate the fermionic action on-shell, and take its derivatives with respect to the corresponding sources. The bulk part of the action actually vanishes on-shell, thus we just have to consider the boundary terms \eqref{bdygravhyp}. As in the bosonic case, this yields a divergent result at $z=0$ and we regularize it at $z=\e$. Direct computation shows that the regularized result contains terms which diverge as $\e^{-2}$ and $\log\e$. The divergences  can be canceled by adding the counterterm action
\begin{align}
\label{bdyct2}
S_\text{c.t.}^\text{F} & =  \int\limits_{z=\e}
\rmd^4x\sqrt{-g}\  \left( \frac 12 \bar\P_m \C^{mrn}  \del_r \P_n 
+ \log \e \left\{- 2 \bar\z \,\dsl\,  \zeta \right. \right. \nn 
& -\frac14  \bar\P_m \C^{mrn} \Box 
\del_r \P_n  + \frac13  \, \f^2 \bar\P_m \C^{mrn} \del_r \P_n \nn
& 
\left. \left.-\frac 16 (\del_n\bar \P_{m}\C^{mn})\dsl(\C^{rs}\del_r\P_s) \right. \right. \nn
& 
\left. \left.  -\frac 23 \rmi
\phi\left[\bar\zeta(\C^{rs}\del_r\P_s)
-(\del_n\bar \P_{m}\C^{mn})\zeta\right]
\right\} \right) \,.
\end{align}
Summing the actions \eqref{bdygravhyp} and \eqref{bdyct2} and sending $\e$ to zero one gets, by construction, a collection of finite terms. After some algebra, using relations \eqref{nearbdychi}--\eqref{nearbdychibeta}  one gets the following result
\begin{align}
\label{univfinite}
& S^\text{F}_\text{ren}  = \lim_{\e\to0}  \left( S^\text{F}_\text{bdy}
+ S_\text{c.t.}^{\text F} \right)  
=  \int \rmd^4 x \left\{\frac 12 (\bar \p^+_{m0}
\tilde\p^-_{m3}+\bar{\tilde\p}^-_{m3}\p^+_{m0}) \right. \nn & \left. -
\bar{\tilde\zeta}^+_1(\zeta^-_0+\rmi a \,\c^-_0 )- (\bar \zeta^-_0 - \rmi
  a \, \bar\c^-_0)\tilde \zeta^+_1+ \frac{\beta a}{2} \,\bar\vartheta^+_0 \,\dsl \,\vartheta^+_0 \right.\nn
&\, \left. -\rmi \beta\bar\vartheta^+_0 ( \zeta^-_0+\rmi a \,  \c^-_0 ) + \rmi \beta
\,  (\bar \zeta^-_0 -\rmi a \bar\c^-_0 ) \vartheta^+_0 + \mbox{scheme-dep}\right\} ~.
\end{align}
Notice that $\theta^+_0$ appears only in terms proportional to $\b$, that is only when supersymmetry is spontaneously broken and the longitudinal component of the gravitino is expected to play a role. Similarly  $\c^-_0$  always appears in combination with $a$, the parameter which controls the breaking of conformal invariance. 

In principle, we would need to solve for the fermionic fluctuations in the bulk to determine how $\tilde\z_1^+$ depends on the sources. However, it is possible to fix such dependence simply considering the transformation properties of the on-shell bulk fields under \eqref{sugratr}. 
We just need the asymptotic solution to \eqref{axgauge}
\begin{equation}
\e(x,z) = z^{-1/2} \e_0^+(x) + z^{1/2} \e_0^-(x) + \mathcal O ( z^{3/2} )\,.
\end{equation}
Under this residual gauge transformation the on-shell bulk fields transform as
\begin{equation}
\d\vth_0^+ = \e_0^+\,, \quad \d\c_0^- = \e_0^-\,, \quad \d \z_0^- = -\rmi a \e_0^- \,,\quad \d \tilde\z_1^+ = -\rmi \b \e_0^+  \ ,
\end{equation}
implying that the following two combinations are invariant
\begin{equation}
\d( \z_0^- +\rmi a \c_0^-)=0 \,,\qquad \d( { \tilde \z}_1^+ +\rmi \b \vth_0^+)=0\ .
\end{equation}
Then, a general solution of the bulk fluctuations is 
\begin{equation}\label{sublead}
{ \tilde \z}_1^+= - \rmi \b \vth_0^+ +  \dsl f(\Box)  ( \z_0^- +\rmi a \c_0^-) + \bar f(\Box) a (\bar \z_0^- -\rmi a \bar\c_0^-)^T\ ,
\end{equation}
with $f$ and $\bar f$ two non-local functions. By substituting the above relation into \eqref{univfinite} one gets
\begin{align}
\label{univfinite2}
S^\text{F}_\text{ren} & =  \int \rmd^4 x \left\{ \frac{\beta a}{2} \,\bar\vartheta^+_0 \,\dsl \,\vartheta^+_0
-2\rmi \beta\bar\vartheta^+_0 ( \zeta^-_0+\rmi a \,  \c^-_0 ) \right. \nn  & \left. + 2\rmi \beta
\,  (\bar \zeta^-_0 -\rmi a \bar\c^-_0 ) \vartheta^+_0 + F(\zeta^-_0,\c^-_0)  + \mbox{scheme-dep}\right\} \,.
\end{align}
where the dependence on $\vartheta^+_0$ is completely fixed. As we will show next, this is all we need in order to derive the supercurrent Ward identities.

In the above action spinors are written in Dirac notation, the natural one in five dimensions, from which \eqref{univfinite2} has been derived. However, since it is in fact a (boundary) 4d action,  it is useful to move from four to two-component spinor notation, the one we used to write down the Ward and operator identities discussed in Section \ref{intro1}. In fact, the boundary leading modes of all bulk fermions, eqs.~\eqref{relfer1} and \eqref{relfer2}, do have a definite chirality (either $+$ or $-$) and hence they source QFT operators with a definite chirality. The precise translation dictionary from a Dirac spinor $\lambda^+$ or $\c^-$ to their Weyl components is
\begin{equation}
\lambda^+ = \lambda_\a \quad,\quad \bar{\lambda}^+ =  \bar{\lambda}_{\dot\a}\quad,\quad \c^- = \bar\c^{\,\dot\a} \quad,\quad \bar\c^- =  \c^\a~.
\end{equation}
This way, the renormalized action \eqref{univfinite2} can be re-written as
\begin{align}
\label{univfinite1}
S^\text{F}_\text{ren} & =  \int \rmd^4 x \left\{ \rmi\frac{\beta a}{2} \,\bar\vartheta_0 \,\dsl \,\vartheta_0
+2\beta\vartheta_0(\rmi \zeta_0 + a \c_0 ) \right. \nn & \left.
- 2 \beta\bar\vartheta_0 (  \rmi\bar\zeta_0 - a \,  \bar\c_0 ) 
   +\dots\right\} \,,
\end{align}
where contraction between Weyl spinors is defined as $\lambda \c = \lambda^\a \c_\a$ and $\bar\lambda \bar\c = \bar\lambda_{\dot\a} \bar\c^{\dot\a}$, and we use conventions where $\l \c=\c\l$ and $(\l\c)^\dagger = \bar\c\bar\l$.

\section{Goldstino from holographic Ward identities}
\label{golhol}

In this section, after spelling-out the exact field/operator map, we show that from the 
renormalized actions \eqref{SrenTfin} and \eqref{univfinite1} all QFT 
Ward identities and operator identities \eqref{wiS}-\eqref{opipsi} can be holographically derived. 

The field/operator map can be read from the 4d linear coupling between the real vector superfield $H^\mu$, where the leading supergravity modes of the graviton multiplet sit, and the FZ multiplet ${\cal J}_\mu$ \cite{Komargodski:2010rb,DiPietro:2014moa}, and that between the hypermultiplet and the QFT operator ${\cal O}$. Upon integration in superspace this reads
\begin{align}
\label{Bbaction}
\int  d^4x & \left[ \frac 12\,  h_0^{\mu\nu} T_{\mu\nu} + \frac 1{2} \left( \rmi \Psi_0^{\mu}  \, S_\mu + c.c. \right) + 2 
\left( \vf_0  {\cal O}_F + c.c. \right) \right. \nn  & \left.-   \sqrt{2} \left(\rmi \zeta_0 {\cal O}_\psi  +  c.c. \right)  + \dots \right] ~,
\end{align}
where the $\dots$ stand for fields we are not presently interested in, as the graviphoton or hyperscalars other than $\f$. The relative normalization between the FZ and ${\cal O}$ multiplets 
is just fixed to get, eventually,  $a$ equal to $\lambda$. From the above action we get the following field/operator map 
\begin{align}
h_0^{\mu\nu}  \longleftrightarrow  \frac 1{2} \; T_{\mu\nu} \quad &, \quad  \Psi_{ 0}^{\mu\a} \longleftrightarrow  \frac \rmi {2} \; S_{\mu\alpha} \\
\vf_0 \longleftrightarrow 2\, {\cal O}_F  \quad &, \quad  \zeta_0^\a  \longleftrightarrow  -  \rmi \sqrt{2} \,{\cal O}_{\psi\,\a} 
\end{align}
Using the decompositions \eqref{comph} and \eqref{gravexp}, the map in the gravitational sector for the operators of interests reads
\begin{equation}
h_0 \longleftrightarrow \frac 12 T \; , \; \vth_0^\a  \longleftrightarrow  -\frac \rmi{2} \; \del^\m S_{\m\a} \; , \;  \bar\c_{0\dot\a} \longleftrightarrow  \frac 1 {2} \; \bar\s^{\m\dot\a\a}S_{\mu\alpha}\ .
\end{equation}
From the action \eqref{SrenTfin} we get
\begin{subequations}
\begin{align}
\label{vT}
\langle T \rangle & = 2 \frac{\delta S^B_{ren}}{\delta h_0} = - 4 \beta a  \\
\label{vOf}
\langle {\cal O}_F \rangle & =  \frac 12 \, \frac{\delta S^B_{ren}}{\delta \vf_0} = - 2 \, \beta 
\end{align}
\end{subequations}
which reproduce the operator identity $\langle T \rangle = 2 \,\l\, {\rm Re}\langle O_F\rangle$, upon the identifcation $a=\l$.
From \eqref{univfinite1} we have
\begin{equation}
\langle \partial^\mu S_{\mu\alpha} (\s^\n\bar{S}_\n)_\b \rangle = - 4 \, \frac{\delta^2 S^F_{ren}}{\delta \vth_{0}^\a \delta \c^\b_0} =  - 8   \beta a \,\ve_{\a\b} ~.
\end{equation}
This, together with \eqref{vT} implies
\begin{equation}
\label{wiShol}
\langle \partial^\mu S_{\mu\alpha} (\s^\n\bar{S}_\n)_\b \rangle =  2 \,\ve_{\a\b} \langle T \rangle~, 
\end{equation}
which exactly reproduces (the $\sigma$-trace of) the QFT Ward identity \eqref{wiS}. From the latter one can derive eq.~\eqref{SS}, i.e. the massless mode associated to the Goldstino. 
This shows that $\langle T\rangle$  is associated to the Goldstino residue in the supercurrent two-point function, as expected for a vacuum with spontaneously broken supersymmetry. Note that from the holographic point of view, this is a complementary, and completely scheme independent way of deriving the VEV of $T$. 
That $\langle T\rangle$ is associated to the Goldstino residue, implies it must be positive, because of unitarity. This in turn discriminates between the different signs of $\beta a$, allowing only for $\b a > 0$ which gives both a positive residue and a positive vacuum energy. Let us finally notice that taking the divergence of \eqref{wiS}, one finds a contact term which is consistently reproduced by the first term in the action \eqref{univfinite1}.

Similarly, one gets from the same action
\begin{equation}
\langle \partial^\mu S_{\mu \alpha} \; {\cal O}_{\psi \beta} \rangle =  \sqrt{2}\rmi \frac{\delta^2 S^F_{ren}}{\delta \vth_0^\a \delta \zeta_0^\b}  =  - 2 \sqrt{2} \, \beta \, \ve_{\alpha\beta}
\end{equation}
which, combined with \eqref{vOf}, exactly reproduces the Ward identity \eqref{wipsi} 
\begin{equation}
\label{wiOhol}
\langle \partial^\mu S_{\mu\alpha} \,  {\CO_\psi}_{\beta} \rangle = \, \sqrt{2} \langle {\cal O}_F \rangle \ve_{\alpha\beta} ~.
\end{equation}
It is worth emphasizing that in order to get eqs.~\eqref{wiShol} and \eqref{wiOhol}, the only information one needs to know about the subleading mode $\tilde\zeta^+_1$ is its local dependence on the source, which we fixed using simple symmetry arguments. In particular, one does not need to know the explicit expression of the non-local function $f$ in  eq.~\eqref{sublead}, which instead depends on the detailed structure of the interior. This is the holographic counterpart of the fact that Ward identities hold in any vacuum and hence independently of the dynamics which generates the VEVs.

Let us finally consider the identity \eqref{opipsi}. Since the sources $\c_0$ and $\z_0$ enter \eqref{univfinite1} and \eqref{sublead} only through the combination $(\z_0-\rmi a\c_0)$, we have
\begin{equation}\label{holoopid}
\frac{\delta S^F_{ren}}{\delta \c_0^\a}  = -\rmi a \frac{\delta S^F_{ren}}{\delta \zeta^{\a}_0} 
\end{equation}
which is nothing but the holographic version of the operator identity $\s^\m\bar{S}_\m = 2\sqrt2 \l \CO_\p$. From the action \eqref{SrenTfin} one can extract similar identities between bosonic operators. Notice, however, that these identities stand on a different footing with respect to the Ward identities \eqref{wiS}-\eqref{wipsi}. The latter contain more dynamical information. In particular, they prove the existence of the Goldstino, which is one of the relevant degrees of freedom of the low energy effective action. This dynamical information cannot be unveiled from an analysis of one-point functions only. 

Further differentiating \eqref{holoopid} with respect to $\bar\z_0$, we find
\begin{equation}\label{lastcorr!}
\langle \s^\mu_{\a\dot\b} \bar S_{\mu}^{\dot\b} \; \bar {\cal O}_{\psi \dot\a} \rangle =  2\sqrt2 \,a\, \langle {\cal O}_{\psi \a} \, \bar {\cal O}_{\psi \dot\a} \rangle\,,
\end{equation}
which gives \eqref{opipsi}. Notice that the correlator \eqref{lastcorr!} should display the massless Goldstino pole. However, in this case this does not arise from contact terms, but rather from the strongly coupled dynamics. Holographically, this means that one would have to solve for the non-trivial fluctuations in the bulk and get the dependence of the subleading modes from the leading ones, which the near-boundary analysis cannot capture.

\section{Conclusions}
\label{concl}

In this work we have provided a holographic description of a general class of supersymmetric quantum field theories in which supersymmetry is spontaneously broken at strong coupling. In particular, by a careful treatment of the holographic renormalization procedure in the fermionic sector, we have recovered a set of Ward identities involving the supercurrent, which encode the presence of the Goldstino, the massless mode associated to the breaking of supersymmetry. 

Our results provide a nice check for the validity of the AdS/CFT duality. In particular, it is rather non-trivial from the bulk side (and consistent with field theory expectations), how the derivation of the Ward identities does not rely on the details of the bulk solution in the deep interior. 
 The approach we used can be applied beyond the class of theories the action \eqref{defqft} and its generalizations describe. Our results provide a powerful tool to distinguish between spontaneous and explicit supersymmetry breaking backgrounds dual to strongly coupled QFTs, independently from issues related to singularity resolution, and in fact from any precise knowledge of the QFT itself.  For instance, following our  strategy, one could inspect several string theory supersymmetry breaking backgrounds proposed in the literature. 

There are several directions one can push further. Our model is a step forward with respect to previous 5d constructions. In particular, differently from the solutions \cite{Gubser:1999pk,Polchinski:2001tt,Argurio:2014rja} used in previous analyses \cite{Argurio:2014rja, Argurio:2012cd,Argurio:2012bi}, our backgrounds break the conformal invariance of the dual SCFT explicitly but in a supersymmetric fashion, like in \cite{Bertolini:2013vka}, a necessary condition for a SQFT to allow for vacua with spontaneously broken  supersymmetry. On the other hand, an important generalization would be to depart from AAdS-ness, and discuss the existence of supersymmetry breaking vacua in more general theories, where the operator responsible for the breaking of conformal invariance is only marginally relevant, like in cascading theories \cite{Klebanov:2000hb}, see \cite{Kuperstein:2014zda} for a recent attempt.

In the same vein, one should consider the issue of (meta)stable dynamical supersymmetry breaking in top-down models directly related to holographic set ups in string theory, like in \cite{Kachru:2002gs,Argurio:2006ny,Argurio:2007qk}. In this perspective, it would be important to find viable non-singular backgrounds, or at least backgrounds where the singularity is as harmless as possible. 

On the other hand, the Goldstino propagator by itself does not probe the stability of the supersymmetry breaking vacuum. In order to say more about vacuum stability  one can try to use holography, and the strategy pursued here, to control the behavior of e.g. the pseudomodulus (the usual suspect as far as tachyonic modes are concerned). Moreover, going beyond the two-point function, one can hope to get a holographic control on the Goldstino effective action \cite{Komargodski:2009rz,Volkov:1973ix} in a strongly coupled setup, in the spirit of \cite{Hoyos:2013gma}.

\section*{Acknowledgements}

We would like to thank Francesco Bigazzi, Ioannis Papadimitriou, Himanshu Raj, Marco Serone, Thomas Van Riet for useful discussions, and Lorenzo Di Pietro for collaboration at the early stage of this project. 
The research of R.A. is supported in part by IISN-Belgium (conventions 4.4511.06, 4.4505.86 and 4.4514.08), by the FWB through the ARC program and by a MIS of the F.R.S.-FNRS. R.A. is a Senior Research Associate of the Fonds de la Recherche Scientifique--F.N.R.S. (Belgium). F.P. acknowledges support by the Netherlands Organization for Scientific Research (NWO) under the VICI grant 680-47-603. This work is part of the D-ITP consortium, a program of the Netherlands Organisation for Scientific Research (NWO) that is funded by the Dutch Ministry of Education, Culture and Science (OCW). The research of D.R. is supported by the ERC Higgs\@ LHC. This work has been supported in part by INFN and COST Action MP1210 The String Theory Universe.


\appendix
\section{The supergravity model: more details}
\label{sugraapp}
The model we consider is $\CN = 2$ 5d supergravity coupled to one hypermultiplet, with scalar 
manifold ${\cal M}=\SU(2,1)/(\U(1)\times\SU(2))$ and the graviphoton gauging a proper $\U(1)$ subgroup of the isometries of ${\cal M}$.  See \cite{Ceresole:2001wi} for a general treatment of this class of theories.

The bosonic field content of the theory includes the metric, one gauge field and four real scalars, $q^X$. The fermionic fields in the theory are the gravitino, $\P_M$, and the hyperino, $\z$, which are both Dirac spinors in our notations. We use capital letter from the middle of the alphabet, $M,\, N$ for curved spacetime indices and capital letters from the beginning of the alphabet, $A,\,B$, for flat indices. Our conventions for 5d Dirac matrices are such that
\begin{equation}
\left\{ \g^A,\, \g^B\right\} = 2 \, \eta^{AB}\,, \qquad \eta=\text{diag}(-1,1,1,1,1)\,.
\end{equation}
The explicit representation we use is
\begin{equation}
\g^\m=\begin{pmatrix}
0	&\rmi\s^\m \\
\rmi\bar\s^\m	&0
\end{pmatrix},\, \m=0,\dots,4\,,\quad \text{and}\quad \g^5=\begin{pmatrix}
1	&0	\\
0	&-1
\end{pmatrix}\,,
\end{equation}
where the $2\times2$ $\s$-matrices are defined as $\s^\m = (-\Id,\, \s^1,\, \s^2,\, \s^3)$. The Dirac conjugate is defined as
\begin{equation}
\bar\P = \rmi \P^\dagger\,\g^0\,.
\end{equation}

The part of the supergravity action which is independent of the gauging can be read from e.g. \cite{Ceresole:2000jd,Sierra:1985ax}. Neglecting cubic and higher terms for spinor and vector fields,  it is
\begin{align}
\label{ungauged}
& S_\text{ungauged}  = \int \rmd^5 x \sqrt{-G} \bigg\{ \,\frac12 R - \bar\P_M\C^{MNP}{\cal D}_N\P_P \nn 
& - 2\bar\z\C^M{\cal D}_M\z  - \frac14  F_{MN}F^{MN} -\frac12 g_{XY} \del_M q^X \del^M q^Y \nn & +  \bar\z \C^M\C^N\P_M \,f_X\,\del_N q^X + \bar\P_M\C^N\C^N\z \,f_X\,\del_N q^X \bigg\}  \,,
\end{align}
where $f_X$ and $g_{XY}$ are functions of the scalar fields and depend on the geometry of the $\s$-model target manifold. In particular, $g_{XY}$ is the metric of a quaternionic manifold parametrised by the four real scalar in the hypermultiplet. Following \cite{Ceresole:2001wi}, we choose this manifold to be $\SU(2,1)/(\U(1)\times\SU(2))$. This is known to be also a K\"ahler manifold and the metric can be derived from the K\"ahler potential
\begin{equation}
\mathcal K = -\frac12 \log\left( S + \bar S - 2 C\bar C \right) \,.
\end{equation}
A convenient parametrization in terms of real coordinates is given by the redefinition
\begin{equation}
S = \ex{2\f} - 1 + \rmi \s\,, \quad C = \tanh(\rho)\, \ex{\f\, + \rmi\a}\,.
\end{equation}
In this coordinate system the quaternionic metric reads
\begin{align}\label{metric}
\rmd q^X \rmd q_X& =2 \left(\rmd\f^2 + \rmd\rho^2\right) + 2 \sinh^2(\rho)\left(\rmd\f^2 + \rmd\a^2\right) \nn & + \frac12 \left(\ex{-2\f} \cosh^2(\rho) \rmd\s +2 \sinh^2(\rho) \rmd\a\right)^2\,.
\end{align}
The full isometry group of the metric above is $\SU(2,1)$ of which we choose to gauge a $\U(1)$ subgroup.\footnote{In the conventions of \cite{Ceresole:2001wi} (see formula (4.19)), the $\U(1)$ gauging corresponds to the choice $\b=-1/3$, $\g=-1/6$.} The gauging procedure, besides promoting the derivatives in \eqref{ungauged}  to their gauge-covariant counterparts, introduces a potential for the scalar fields as well as interaction terms for the fermions. As anticipated in the main text, we are here interested in single-scalar backgrounds. We thus simplify our model fixing $\s = \rho = \a = 0$. The gauged action, truncated to the desired field content and neglecting four-fermion interactions, reads
\begin{align}
& S_\text{5D}  = \int \rmd^5 x \sqrt{-G} \bigg\{ \,\frac12 R -  \bar\P_M\C^{MNP}D_N\P_P - 2\, \bar\z\C^M D_M\z \nn 
& -\del_M \f\, \del^M \f 	- U(\f)
+ \rmi \del_N \f \left(  \bar\z \C^M\C^N\P_M \,  - \bar\P_M\C^N\C^M\z
\right)  \nn &
+ 2\,\CN(\f) \left( \bar\P_M\C^M\z + \bar\z\C^M\P_M
\right) + m(\f)\, \bar\P_M \C^{MN}\P_N \nn & - 2\,\CM(\f)\, \bar\z\z  \bigg\} \,, 
\label{5daction}
\end{align}
where we also neglect the terms containing the graviphoton, since they will play no role in the following discussion. 
The derivatives are standard space-time covariant derivatives. When acting on a spinor they read
\begin{equation}
 D_M = \del_M + \tfrac14 \o_{M AB}\g^{AB} \,,
\end{equation}
with $\o$ the space-time spin connection.
The scalar potential and superpotential are given by
\begin{align}
\label{potsup}
&U(\f) = \frac1{12} \left( 10 - \cosh(2\f) \right)^2 - \frac{51}4  \,, \nn
&W(\f) = \frac16 \left( 5 + \cosh(2\f) \right)
\end{align}
where the former is related to the latter through the equation
\begin{equation}
\label{reluw}
 U =   \frac94 \, \del_\f W\del_\f W -6  W^2\,.
\end{equation}
The other quantities which enter \eqref{5daction} are given by
\begin{subequations}
\label{fermmass}
\begin{align}
m(\f) & = \frac32 W(\f) = \frac14 \left( 5 + \cosh(2\f) \right)		\,,	\\
\CM(\f) & = \frac92 W(\f) -5 = -\frac14 \left(  5- 3\cosh(2\f) \right)	\,,	\\
\CN(\f) & = -\frac34\rmi \del_\f W(\f) = -\frac\rmi4 \sinh(2\f)		\,.
\end{align}
\end{subequations}
\section{Boundary terms for gravitino and hyperino}
\label{bdyferm}
In this appendix we briefly outline how to obtain the fermionic boundary action, following the procedure given for instance in \cite{Henneaux:1998ch}.

Let us start from the action \eqref{Lferm2}, written for simplicity in a pure AdS background since it will become clear that the boundary term will not depend on the non-trivial bulk profiles
\begin{align}
S^\text{F}_\text{5D} &= \int \rmd^5 x \sqrt{-G} \bigg\{ -  \bar\P_M\C^{MNP}D_N\P_P - 2\, \bar\z\C^M D_M\z \nn &
 - \frac32\, \bar\P_M \C^{MN}\P_N + \bar\z\z  \bigg\} \,.
\label{Lferm2ads}
\end{align}
The masses actually are there just to instruct us on what is the behaviour in $z$ and the chirality of the leading fermionic modes near the boundary
\begin{equation}
\P_m = \P_{m0}^+ z^{-1/2}+\dots, \qquad \z = \z_0^- z^{3/2} + \dots
\label{bdychir}
\end{equation}
Up to a boundary term, the action \eqref{Lferm2ads} can be recast into an explicitly real expression
\begin{align}
S^\text{F}_\text{5D} & = \int \rmd^5 x \sqrt{-G} \bigg\{ - \frac12 \bar\P_M\C^{MNP}D_N\P_P \nn &
+\frac12 D_N \bar\P_M\C^{MNP}\P_P 
-  \bar\z\C^M D_M\z +D_M\bar\z\C^M \z+\dots  \bigg\} \,.
\label{Lferm2real}
\end{align}
We now take the variation of the above action, keeping the leading modes fixed at the boundary.  In other words, the (leading) variations of the gravitino and hyperino will be of the opposite chiralities with respect to \eqref{bdychir}. On shell the variation of \eqref{Lferm2real} still yields a non-trivial boundary term
\begin{align}
\delta S^\text{F}_\text{5D} &= -\int \rmd^4 x \sqrt{-g} \bigg\{- \frac12 \bar\P_m^+\C^{mn}\delta\P_n^-
-\frac12 \delta \bar\P_m^-\C^{mn}\P_n^+ \nn &
-  \bar\z^- \delta\z^+ -\delta\bar\z^+ \z^-  \bigg\}\, .  
\label{Lferm2var}
\end{align}
In order for the action to be stationary on the bulk equations of motion, we need to supplement it with a boundary term whose variation exactly cancels the one above. Such a boundary term is the following
\begin{align}
S^\text{F}_\text{bdy} = \int \rmd^4 x \sqrt{-g} \left\{ - \frac12 \bar\P_m \C^{mn} \P_n - \bar\z\z \right\}\,.
\end{align}
It is then straightforward to see that the above boundary term coincides with the on-shell action, since the bulk part \eqref{Lferm2real} exactly vanishes on the fermionic equations of motion.

We remark that the end result above does not agree with some previous attempts in the literature for a massive gravitino (see for instance \cite{Rashkov:1999ji}).


\bibliography{biblio_gold}

\end{document}